\documentstyle[12pt]{article}
\begin{document}
\begin{center}
\baselineskip 0.80cm
\vskip 0.0in
{\Large \bf 
Orthorhombicity mixing of s- and d- gap\\
components
in $YBa_2Cu_3O_7$\\
without involving the chains}
\vskip 0.9in
{\large \bf
G. Varelogiannis} 
\vskip 0.04in
\vskip 0.01cm
{\em Institute of Electronic Structure and Laser\\      
Foundation for Research and Technology - Hellas\\
P.O. Box 1527, Heraklion Crete 71110, Greece}
\vskip 0.21in          
\vskip 0.1in
\begin{abstract}
\baselineskip 0.80cm
Momentum decoupling develops when forward scattering dominates the
pairing interaction and implies tendency for decorrelation between
the physical behavior in the various regions of the Fermi surface.
In this regime
it is possible to obtain anisotropic s- or d-wave superconductivity even
with isotropic pairing scattering.
We show that in the momentum decoupling regime the
distortion of the $CuO_2$ planes is enough
to explain the experimental
reports for s- mixing in the dominantly d-wave gap of $YBa_2Cu_3O_7$.
In the
case of spin fluctuations mediated pairing instead,
a large part of the condensate
must be located in the
chains in order to understand the experiments.
\end{abstract}
\vskip 0.9cm
\vskip 0.2cm
PACS numbers: 74.25.-q \hfill
\end{center}
\newpage
\baselineskip 0.80cm

The issue of the symmetry of the order 
parameter in the oxides motivated intense 
investigations \cite{review}.
Advanced phase sensitive experiments have been developed recently that
allowed to establish that the order parameter in $YBa_2Cu_3O_7$
reverses its sign on the Fermi surface indicating d-wave symmetry
\cite{phaseD}.
This symmetry is generally believed to indicate spin fluctuation
mediated superconductivity.
The presence of nodes in the gap of $YBa_2Cu_3O_7$
is confirmed by the linear temperature
dependence of the penetration depth in the low temperature
regime \cite{nodesD}. However, there are also results that
are in clear conflict with a simple d-wave picture
\cite{Chaudhari}.
In particular, c-axis Josephson tunneling experiments on $YBa_2Cu_3O_7$
indicated the existence of a significant s-component
\cite{DynesPb}.
This late conclusion is reinforced by the relative insensitivity
of the superconducting critical temperature on the presence of non-magnetic
impurities or defects \cite{DynesImp}. 
It appears experimentally that the gap has
a dominant d-wave component but also a significant s-wave
component. It has been argued that this behavior may indicate 
the existence of two different condensates \cite{Muller}.

The mixing of s and d components arises naturally when the lattice
is orthorhombically distorted
\cite{James}. Large orthorhombic
distortions have therefore been invoked in order to understand the
experimental conflicts in $YBa_2Cu_3O_7$ \cite{Maki,Pokrovsky,Jules}.
However, the orthorhombic distortion of the $CuO_2$ planes 
in the case of $YBa_2Cu_3O_7$ is only a few percent ($\approx 3\%$) 
and such a small distortion cannot induce significant mixing of 
s-components in a d-wave spin fluctuations mediated pairing.
To reconcile the 
large orthorhombicity effects required by the phenomenology and
spin fluctuations pairing,
it has been argued that the $Cu-O$ chains 
are involved in superconductivity
and at least $25\%$ of the condensate is located there \cite{Jules}.
Since the chain band concerns only one direction in 
the $ab$ plane, if chains are involved, large in plane anisotropies
are reasonable. Large anisotropies between the $a$ and $b$
directions are also reported in microwave penetration depth
measurements \cite{penab} and in dc resistivity measurements
\cite{Friedmann}. On the other hand, supposing that the chains
contain a large part of the condensate and are therefore crucially
involved in the pairing mechanism is difficult to reconcile with 
the fundamental similarities of superconductivity in $YBCO$ with
that of the other high-$T_c$ cuprates where the chains are absent. 
Whether the chains are involved in the pairing or not
is not yet a definitely answered question,
there are nevertheless strong arguments supporting that only
$CuO_2$ planes are involved in the interesting physics \cite{PWA}.

It has been proposed recently 
an alternative to spin fluctuations mechanism of 
anisotropies and gap symmetry transitions that
involves isotropic scattering and has been named Momentum 
Decoupling (MD) \cite{meMD1,meMD2,meMD3}. 
When the characteristic momenta exchanged in the pairing interaction
are small compared to the characteristic momenta of the variations
of the electronic density of states, there is tendency for decorrelation 
between the physical behavior in the different regions of the Fermi surface.
In particular, couplings become proportional to the
angular resolved electronic density of states (ARDOS) $N(E_F,\vec{k})=
|\upsilon_F(\vec{k})|^{-1}$
at each region of the Fermi surface, and therefore anisotropies are 
driven by the electronic density of states and not by the 
scattering \cite{meMD1}. Taking into account the conventional Coulomb 
pseudopotential $\mu^*$ the d-wave and s-wave 
(both ARDOS driven anisotropic) states become energetically degenerate 
\cite{meMD2}.
The presence of different gap symmetries in different oxides as well 
the d-s gap symmetry transition by overdoping
$Bi_2Sr_2CaCu_2O_8$ \cite{Kelley}            
are natural consequences of MD \cite{meMD2,meMD3}.
The temperature enhancement of the anisotropy \cite{Ma}
and the behavior of the anomalous dip above the gap in the electronic
density of states
\cite{meDIP} are qualitative puzzling aspects of the 
phenomenology of $Bi_2Sr_2CaCu_2O_8$ that also indicate MD \cite{meMD1}.

Dominance of forward scattering in the pairing
could result from the vicinity of the strongly correlated 
electronic system to a 
phase separation instability that could be driven by
magnetic fluctuations \cite{Marder} or even by phonons                    
\cite{Hubb}.
The interlayer tunneling mechanism proposed by Anderson
is effectively $q\approx 0$ pairing and could 
be at the origin of MD \cite{PWAtun}.
The same for the charge transfer resonance pairing mechanism
\cite{CTR} which also concerns small 
momentum transfer process \cite{Littlewood}.
Notice that dominantly forward scattering has unexpected 
implications even for the normal state properties
that have not been yet fully explored like for example the possibility
of linear $T$-dependent dc resistivity despite electron                      
scattering with high energy phonons \cite{dc}.
We report here that
the  
orthorhombic distortion of the $CuO_2$ planes in $YBa_2Cu_3O_7$,
produces an effect an order of magnitude larger in the case
of MD than in the
case of spin fluctuations pairing
and could, therefore,
explain the experimental reports of significant 
mixing of s- components in the dominantly d-wave gap without need to involve
the chains.          

We solve the BCS equations on a two dimensional
lattice that might simulate the $CuO_2$ planes of $YBCO$. The gap is obtained by
$$
\Delta(\vec{k})=-\sum_{\vec{p},|\xi_{\vec{p}}|<\Omega_D}
{\Lambda(\vec{k}-\vec{p})\Delta(\vec{p})\over
2\sqrt{\xi_{\vec{p}}^2+\Delta(\vec{p})^2}}
\tanh\biggl(
\sqrt{{\xi_{\vec{p}}^2+\Delta(\vec{p})^2\over 2T}}\biggr)
\eqno(1)
$$
The materials characteristics enter through the dispersion $\xi_{\vec{k}}$.
The effect of orthorhombicity on the $CuO_2$ plane is to make 
inequivalent the $a$ and $b$ axes and in $YBa_2Cu_3O_7$ the
difference in these lattice constants is less than $\approx 3.5\%$. 
For such small variations we can consider  
that in a tight-binding dispersion the hoping
along the two different axes will be inequivalent
with differences of the same 
order. 
We consider in fact a simple next nearest neighbors tight binding fit to LDA
calculations of the $CuO_2$ band in $YBCO$ \cite{OKAnd}
$$
\xi_{\vec{k}}=-2t [\cos(k_x) + (1+\beta)\cos(k_y)]-4t'
\cos(k_x)\cos(k_y) - \mu 
\eqno(2)
$$
where
$t=0.25eV$, $t'/t=-0.45$ and $\mu=-0.44 eV$. This type of dispersion 
produces a van Hove peak in the density of states about $10 meV$ below the 
Fermi level. The relevant parameter for our discussion is $\beta$ which
characterizes the orthorhombic distortion.                            

The scattering amplitude $\Lambda(\vec{k}-\vec{p})$ in equation (1) contains the
physics of the pairing mechanism. The two different situations 
of Momentum Decoupling and spin fluctuation pairing that we consider here 
correspond to two different characteristic structures of $\Lambda(\vec{k}-\vec{p})$.
In the momentum decoupling regime the pairing scattering is isotropic taking 
at small momenta a lorentzian form
$$
\Lambda(\vec{k}-\vec{p})=-\Lambda^o\biggl( 
1+{|\vec{k}-\vec{p}|^2\over q_c^2}\biggr)^{-1}+\mu^*
\eqno(3)
$$
where the first term concerns the pairing and
$q_c$ plays the role of a momentum cutoff. This type of lorentzian
form is found to occur in the scattering of the electronic system with any
bosonic system including phonons, 
provided the electronic system is close to the phase 
separation instability \cite{Hubb}. The Coulomb pseudopotential $\mu^*$ is
the effective repulsion of the
paired electrons and is not necessarily momentum independent. We are in the MD
regime provided the characteristic momenta of the variations of $\mu^*$ are large
compared to $q_c$. 

The interaction of equation (3) leads to either s- or 
d-wave superconductivity, depending on marginal for the pairing parameters like
the magnitude of $\mu^*$ and its characteristic momentum range. 
Considering for
$\mu^*$ a Lorentzian structure as that of the pairing amplitude
we were able to plot a phase diagram of the energetically 
favorable (having the lowest free energy) 
gap symmetry (s-wave or d-wave) on a plane defined by
the ratio of the characteristic cut-off of $\mu^*$ over that of 
the pairing amplitude and the magnitude of $\mu^*$
for an electronic structure similar to that of the oxides \cite{meMD2,meMD3}.
What is relevant for our discussion here is that a dominantly d-wave gap
as reported by phase sensitive and node 
sensitive experiments on $YBCO$ arises naturally for conventional values of $\mu^*$
with a pairing amplitude as
in equation (3) \cite{meMD2,meMD3}. 

The alternative ``conventional'' mechanism for d-waves is the 
scattering with spin fluctuations that has been extensively discussed in 
the literature.
As an example of this second approach 
we consider the phenomenological Millis Monien and Pines (MMP)
scattering with spin fluctuations \cite{MMP} in the static limit
$$
\Lambda(\vec{k}-\vec{p})\approx 
{-\Lambda_o\over 1 + \xi^2_M(\vec{k}-\vec{p}-\vec{Q})^2}
\eqno(4)
$$
where $\vec{Q}=(\pi,\pi)$, the coherence range of the 
antiferromagnetic
spin fluctuations $\xi_M$ is taken
on the order of three lattice spacings as in the 
experiment \cite{MMP} and Coulomb pseudopotential is neglected.

In the orthorhombically distorted case $a$ and $b$ directions are not 
equivalent and since the Fermi velocities are different in these 
two directions one would expect different magnitudes of gap.
The difference between the absolute values
of the gap along $a$ and along $b$ is therefore a measure of the
orthorhombicity effect. We plot in figure (1a) the evolution
of the ratio $\Delta_a^2/\Delta_b^2$ with $\beta$.
In the tetragonal case $\beta=0$ 
this ratio is of course equal to unity. However, as 
we switch on the distortion $\beta$ the maximum absolute values 
of the gap we obtain near the $(0,\pi)$ and $(\pi,0)$ points are 
appreciably different. Full line in figure 1a corresponds to the 
MD regime with a scattering amplitude as in equation (3) and
dashed line to the MMP scattering amplitude given in Eq. (4).
In both cases the energetically favorable d-wave channel is considered
and therefore the gap changes sign between $(0,\pi)$ and $(\pi,0)$.
We can already conclude from figure 1a that in the case of MD the effect 
of orthorhombicity is an order of magnitude larger than in the case of
spin fluctuations. 

Let us illustrate now that, in the MD case, the distortion 
of the $CuO_2$ planes may be sufficient to understand the experiments.
We first consider the London penetration 
depth along the two different directions at zero temperature
$$
\lambda_{k_x(k_y)}^{-2}\propto \sum_{\vec{k}}\upsilon_{k_x(k_y)}^2
\biggl(
\partial f (E_{\vec{k}})
/ \partial E_{\vec{k}}\biggr)
\eqno(5)
$$
where $E_{\vec{k}}=\sqrt{\xi_{\vec{k}}^2+\Delta_{\vec{k}}^2}$.
The experimental results of Ref. \cite{penab} indicate large in-plane
anisotropy of the penetration depth $\lambda_a/\lambda_b\approx 1.6$.
We show in figure (1b) the dependence of the 
penetration depth in plane anisotropy $\lambda^{-2}_a/\lambda^{-2}_b$
on the distortion parameter $\beta$. The full line corresponds to 
the MD regime while the dashed line to the MMP spin-fluctuation
scattering. We see that in the MD regime the in plane distortion
expected on the order $\beta\approx 0.3-0.4$
could be sufficient to produce the experimental in-plane anisotropy 
of the penetration depth, while for an MMP interaction, the 
reported in plane anisotropy of $\lambda$ is an order of magnitude
smaller than in the experiment. 

The same can be said
for the c-axis Josephson tunneling results  
of Dynes and collaborators \cite{DynesPb}. In fact they observed 
Josephson tunneling currents on
c-axis $Pb$/insulator/$YBa_2Cu_3O_7$ 
tunnel junctions. According to Ambegaokar and Baratoff \cite{Ambegaokar}
the Josephson current is given by
$$
JR={2\pi T \over 
N_1N_2}{1\over \pi}
\sum_{n=0}^{\infty}
\sum_{\vec{k}}{\Delta_1(\vec{k})\over \xi_1(\vec{k})^2+\Delta_1(\vec{k})^2+
\omega_n^2}
\sum_{\vec{k'}}
{\Delta_2(\vec{k'})\over \xi_2(\vec{k'})^2+\Delta_2(\vec{k'})^2+
\omega_n^2}
\eqno(6)
$$
At zero temperature the sum over the fermion Matsubara frequencies
is becoming an integral that can be performed straightforwardly,
leading to the following expression for the Josephson current
at $T=0$:
$$
J(T=0)R={1\over 2\pi}{1\over N_1N_2}
\sum_{\vec{k}\vec{k'}}\Delta_1(\vec{k})\Delta_2(\vec{k'})
{1\over
\sqrt{\xi_1(\vec{k})^2+\Delta_1(\vec{k})^2}
\sqrt{\xi_2(\vec{k'})^2+\Delta_2(\vec{k'})^2}}
\times
$$
$$
\times
{1\over
\sqrt{\xi_1(\vec{k})^2+\Delta_1(\vec{k})^2}+
\sqrt{\xi_2(\vec{k'})^2+\Delta_2(\vec{k'})^2}}
\eqno(7)
$$
where $R$ is the junction resistance and $N_i(0)$ the densities 
of states on the Fermi level.                 
It is clear that if $\Delta_1$ and $\Delta_2$ are orthogonal
(they belong to different irreducible representations of
the point group),
there should not be any Josephson current in the junction.
Therefore, since the gap of $Pb$ is known to be s-wave,
the observation of the Josephson current seems to exclude
a purely d-wave gap in $YBCO$ and a significant part of s-component 
is necessary in order to have Josephson coupling between the 
two condensates.                       
For the $Pb$/insulator/$YBCO$ junction, if we suppose that the $Pb$
gap is isotropic then in equation (6) the sum over $k$ for the 
isotropic case is becoming trivial leading to a term proportional to 
to the density of states of lead. At 
zero temperature the matsubara frequency sum is
becoming a
frequency integral taking here the form
$\int_0^\infty d\omega F(\omega)G(\omega)$ where
$F(\omega)=(\Delta_{Pb}^2+\omega^2)^{-1/2}$ and $G(\omega)=
(\xi_Y(\vec{k})^2+\Delta_Y(\vec{k})^2+\omega^2)^{-1}$. 
This integral is calculated numerically.

In Ref. \cite{DynesPb} is reported a Josephson
current along the c axis that was about $10\%$ of what it should be   
expected from the isotropic Ambegaokar-Baratoff formula \cite{Ambegaokar} if for 
YBCO the gap were taken equal to $1.76 T_c$ as expected in
weak coupling BCS theory. 
The weakness of the supercurrent could show that the d-components
are dominant in $YBCO$ \cite{Clemm}. 
To our approach the gap in $YBCO$ is indeed dominantly d-wave
yet because of the
orthorhombic distortion there is also an s-component that is responsible for the 
Josephson coupling with the condensate of lead.
To show that this approach could reasonably account for the results 
of \cite{DynesPb} we take two different cases. In the first case
we consider that the gap of $YBCO$ is isotropic
and in the second case we obtain the gap from the solution of the
BCS equations as previously. In both cases we adjust the $YBCO$ gap
to a value about $15$ times larger than the gap of $Pb$. We also
adjust the isotropic gap we take for YBCO in the first case to be
equal to $(1/2)(|\Delta_a|+|\Delta_b|)$.
What would be comparable to the findings of Ref. \cite{DynesPb}
is the ratio of the Josephson current that results
using the anisotropic gap we 
obtain in the MD regime solving                  
the BCS equations as previously over the supercurrent
obtained in the isotropic case and which should correspond to the 
Ambegaokar-Baratoff expectations. We plot in figure (1c)
the evolution of this ratio with the distortion parameter $\beta$.
When $\beta = 0$ we have no Josephson supercurrent 
and as the distortion parameter reaches values as high as 
$\beta=0.04$ in the case of MD (full line) 
we can have appreciable supercurrents of the order
of $15\%$ of what should be expected in a junction between 
isotropic superconductors in agreement with the results of \cite{DynesPb}.
With the MMP interaction instead the supercurrent is here also 
about an order of magnitude smaller than the experimental report.

It emerges therefore a fundamental qualitative difference
between MD and spin fluctuations pairing. 
In the later case,                           
if the orthorhombic distortions interpretation of the 
$s$ and $d$ mixing in $YBCO$ makes sense, the chains participate
fundamentally in the pairing and at least 
about $25\%$ of the condensate should be 
located there. On the other hand, in the case of MD, the orthorhombic
distortion of the $CuO_2$ planes is sufficient. 
We proposed therefore a mechanism that explains the puzzle of 
significant s-wave components in $YBa_2Cu_3O_7$ without contradicting  
the strong arguments \cite{PWA} supporting that the relevant physics happens in 
the $CuO_2$ planes.

Discussions with J.F. Annett and
E.N. Economou are gratefully acknowledged.

\newpage

{\Large\bf Figure Captions:}

\vskip 0.3cm

{\bf Figure 1:} (a): the ratio of the gaps along the $a$ and $b$
directions $\Delta_a^2/\Delta_b^2$ as a function of the distortion
parameter $\beta$. (b): The London penetration depth in-plane anisotropy
$\lambda_b^2/\lambda_a^2$ as a function of the distortion parameter
$\beta$. (c): The ratio of supercurrent obtained from
a Josephson junction of $Pb$ with anisotropic $YBCO$
over that expected from a junction of lead with isotropic $YBCO$ with gap
magnitude $(1/2)(|\Delta_a|+|\Delta_b|)$. In all cases the full lines
correspond to the MD regime as described in the text and dashed lines
to the MMP spin fluctuations scattering amplitude with 
the same dispersion conditions. 

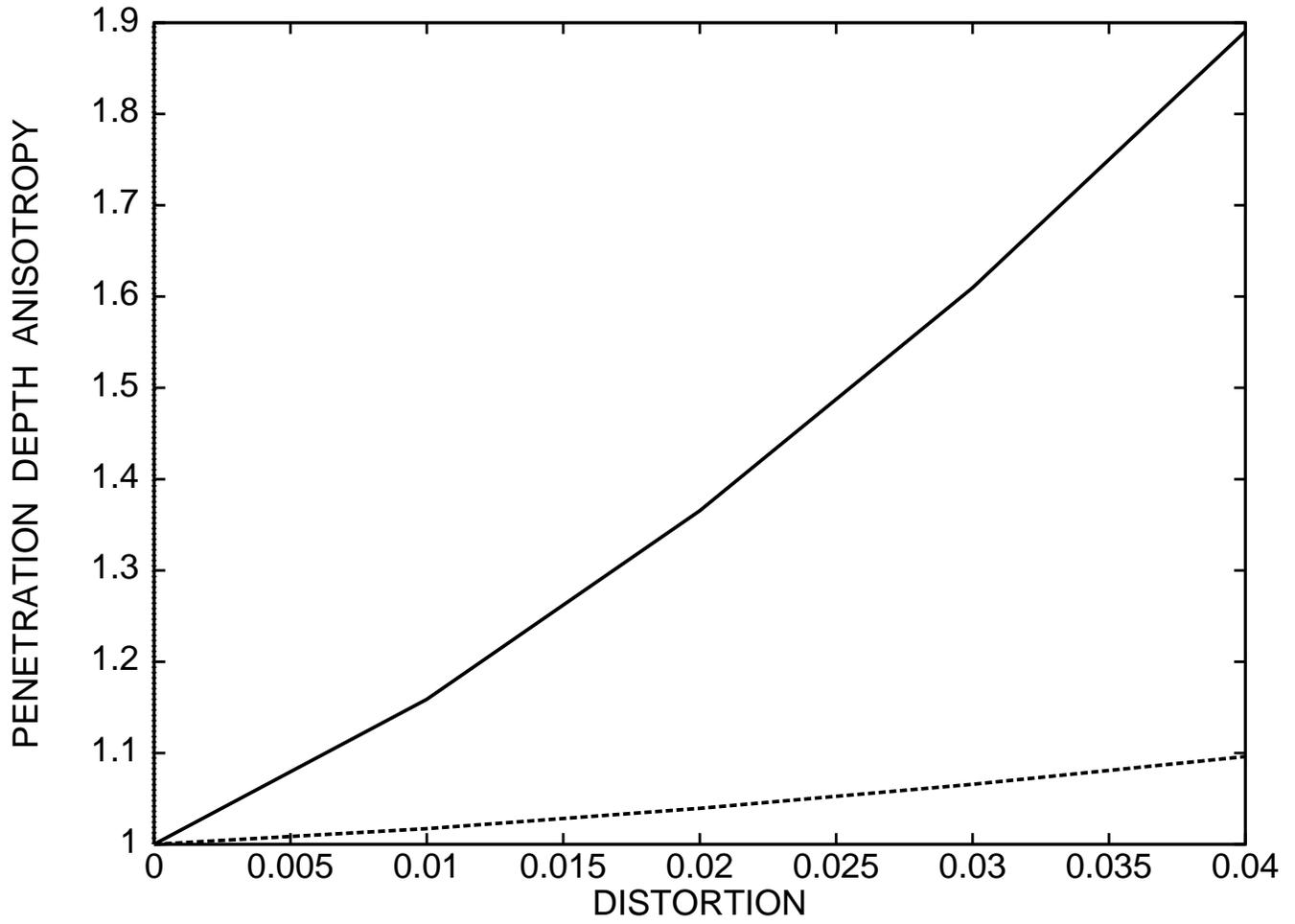
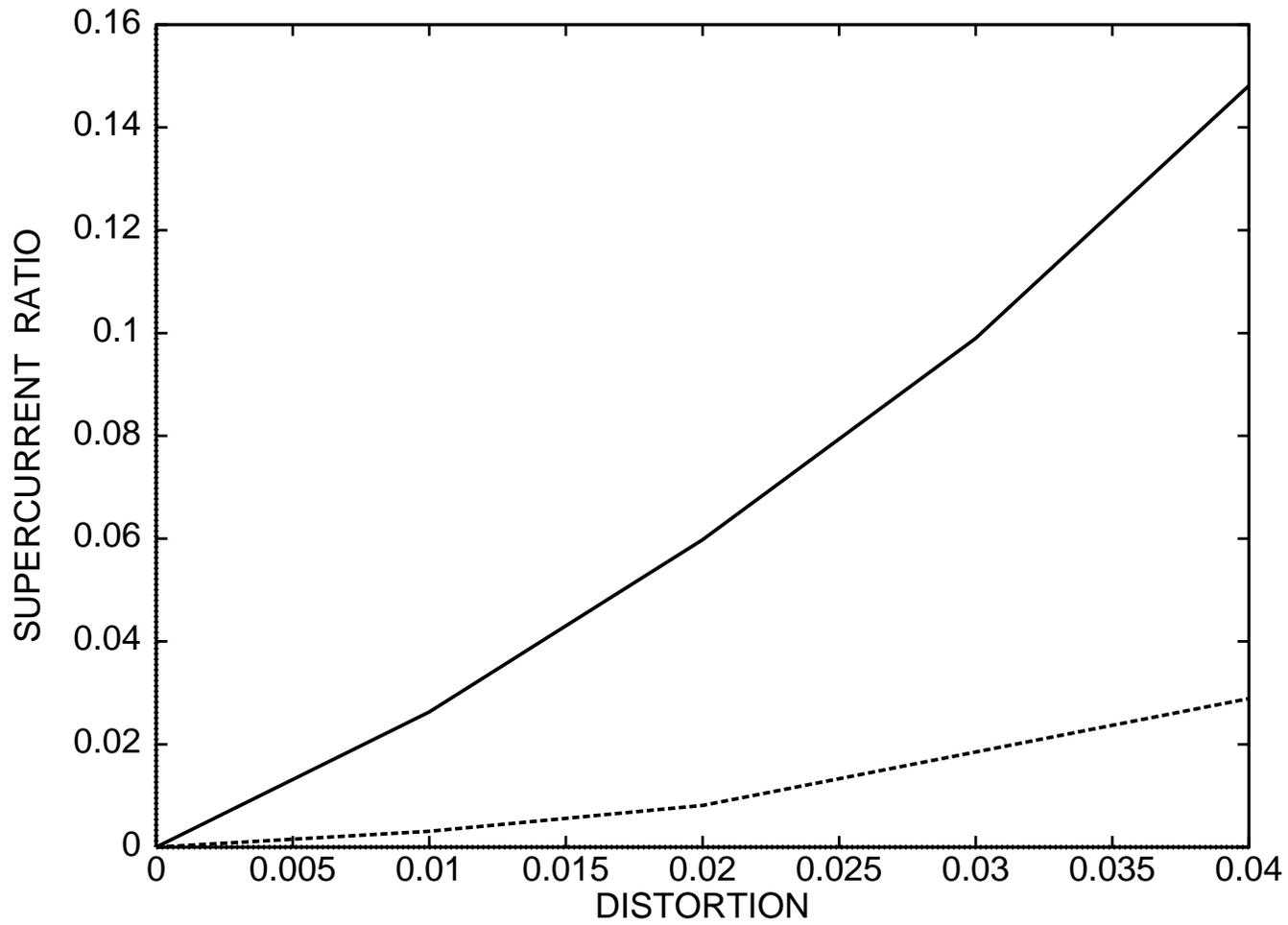
\end{document}